\documentclass[]{spie}  %>>> use for US letter paper
\usepackage[]{graphicx}
\usepackage{wrapfig}

\title{Twilight observations with MASS-DIMM} 

\author{Boris S. Safonov\supit{s}
\skiplinehalf
\supit{s}Sternberg Astronomical Institute, Universitetsky pr-t, 13, Moscow, Russia, 119992
}

\authorinfo{Further author information: e-mail: safonov@sai.msu.ru}
\begin{document} 
  \maketitle 

%%%%%%%%%%%%%%%%%%%%%%%%%%%%%%%%%%%%%%%%%%%%%%%%%%%%%%%%%%%%% 
\begin{abstract}
In this paper we describe the method of measurement of optical turbulence in twilight with MASS-DIMM. Some results of such observations carried out with SAI ASM installed on Mt. Shatdzatmaz: 1) characteristics of optical turbulence don't change significantly in transition between day and night 2) twilight measurements of seeing can be used for selection of observational program for big telescope on condition that program switching is long --- longer than 30-40 minutes. In conclusions we list upgrades of ASM system needed to improve capabilities to observe on twilight.
\end{abstract}


%%%%%%%%%%%%%%%%%%%%%%%%%%%%%%%%%%%%%%%%%%%%%%%%%%%%%%%%%%%%%
\section{INTRODUCTION} \label{sec:intro}

Normally MASS-DIMM is used in night for evaluation of optical turbulence in usual time of astronomical observations. But observing in twilight might be interesting either. Els at al (2009)\cite{Els09} pointed to the fact that OT layers strenghts alter during night. Hence it's expectable that in transition between day and night and vice versa changes can be even greater. Besides evening twilight observations can be used for selection of observational program for big telescope for forthcoming night. The same valid for morning twilight observation what can be used for solar observations.

\begin{figure}[b]
	\begin{center}
	\begin{tabular}{cc}
	\includegraphics[height=6.5cm]{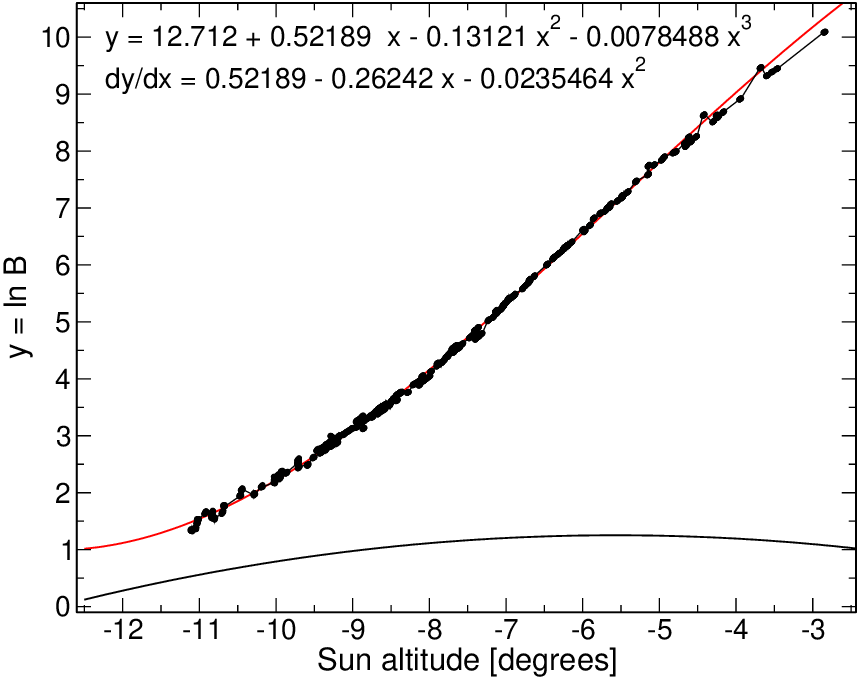}
	\includegraphics[height=6.6cm]{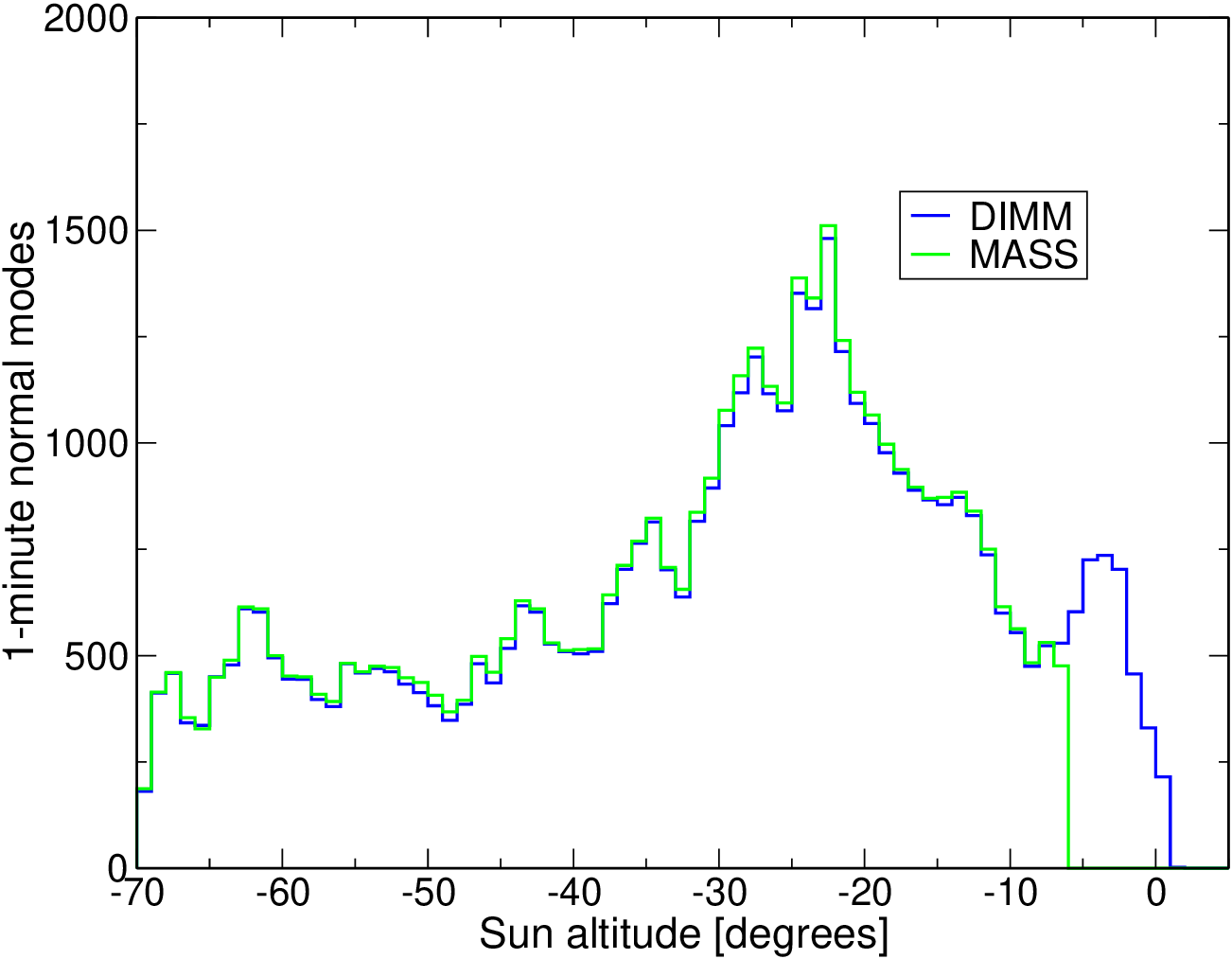}
	\end{tabular}
	\end{center}
	\caption
	{\label{fig:flux_mass_sunalt_distr}
	{\it Left:} star flux in MASS with polynomial approximation. {\it Right:} Distribution of observational time in period from November 09 to August,10.
	}
\end{figure}

To test this suggestions we observed with SAI ASM station installed on mt. Shatdzatmaz\cite{Kornilov10}. We made some modification of algorithm of observations in order to allow for twilight measurements: 1) MASS measures at $h_{\odot}<-6^{\circ}$ (see sec. \ref{sec:method:mass}), DIMM --- at $h_{\odot}<0^{\circ}$ (see sec. \ref{sec:method:dimm}) 2) during twilight background measurements cadence was increased (see sec. \ref{sec:method:mass}). Twilight observations were started in automatic mode in November of 2009; Distribution of observational time in period from November'09 to August'10 is represented on figure \ref{fig:flux_mass_sunalt_distr}, right. Table \ref{tab:twilight} respresents some statistics of twilight observations. 

Next section will be about some differencies between nighttime and twilight measurements with MASS-DIMM (sec. \ref{sec:method}). Then we will discuss how status of turbulence above our particular summit depends on sun altitude in twilight (sec. \ref{sec:ot}). After this we will analyse possibility of prediction of seeing for forthcoming night (sec. \ref{sec:prediction}). Conclusions of this work are given in section \ref{sec:conclusions}.

\section{PECULIARITIES OF TWILIGHT OBSERVATIONS WITH MASS-DIMM} \label{sec:method}

Our ability to observe in twilight is limited by the level of sky background because it reaches high values and changes rapidly. Let's consider how background affects observations with MASS-DIMM. ASM includes 3 devices sensitive to background: guider, DIMM and MASS.

\begin{table}[h]
\caption{\footnotesize Total duration of twilight measurement.}\label{tab:twilight}
%\centering
\footnotesize
\begin{center}
\begin{tabular}{l|c|cc|cc|l}
 & & \multicolumn{2}{c|}{DIMM} & \multicolumn{2}{c}{MASS} & \\
twilight & $h_{\odot}$ limits & evening & morning & evening & morning & units\\
\hline
total &   & 60 & 58 & & & days \\
civil & $(-6^{\circ},0^{\circ})$ & 24.9 & 38.0 & & & hours\\
nautical & $(-12^{\circ},-6^{\circ})$ & 24.8 & 32.1 & 24.4 & 32.5 & hours\\
astronomical & $(-18^{\circ},-12^{\circ})$ & 39.1 & 48.3 & 39.5 & 48.9 & hours\\
\end{tabular}
\end{center}
\end{table}

\begin{figure}
	\begin{center}
	\begin{tabular}{cc}
	\includegraphics[height=7cm]{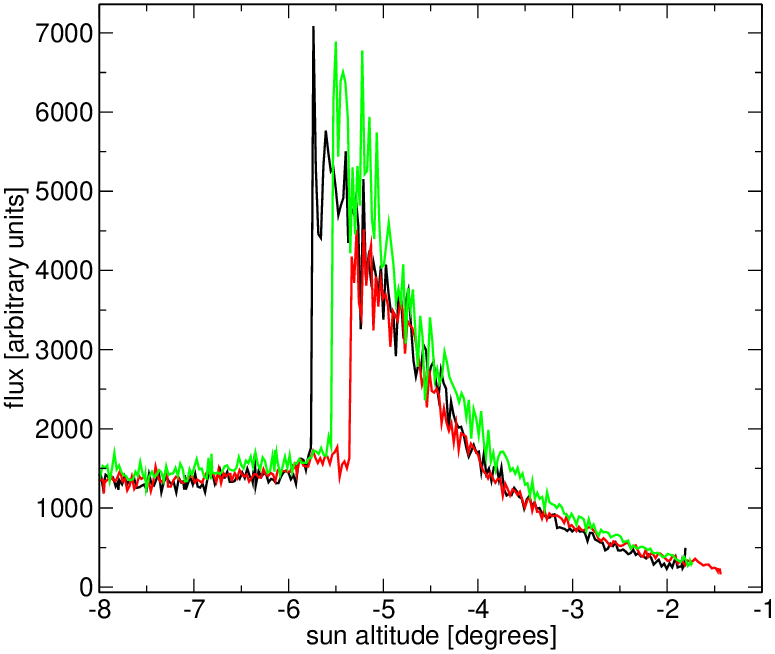}
	\includegraphics[height=7cm]{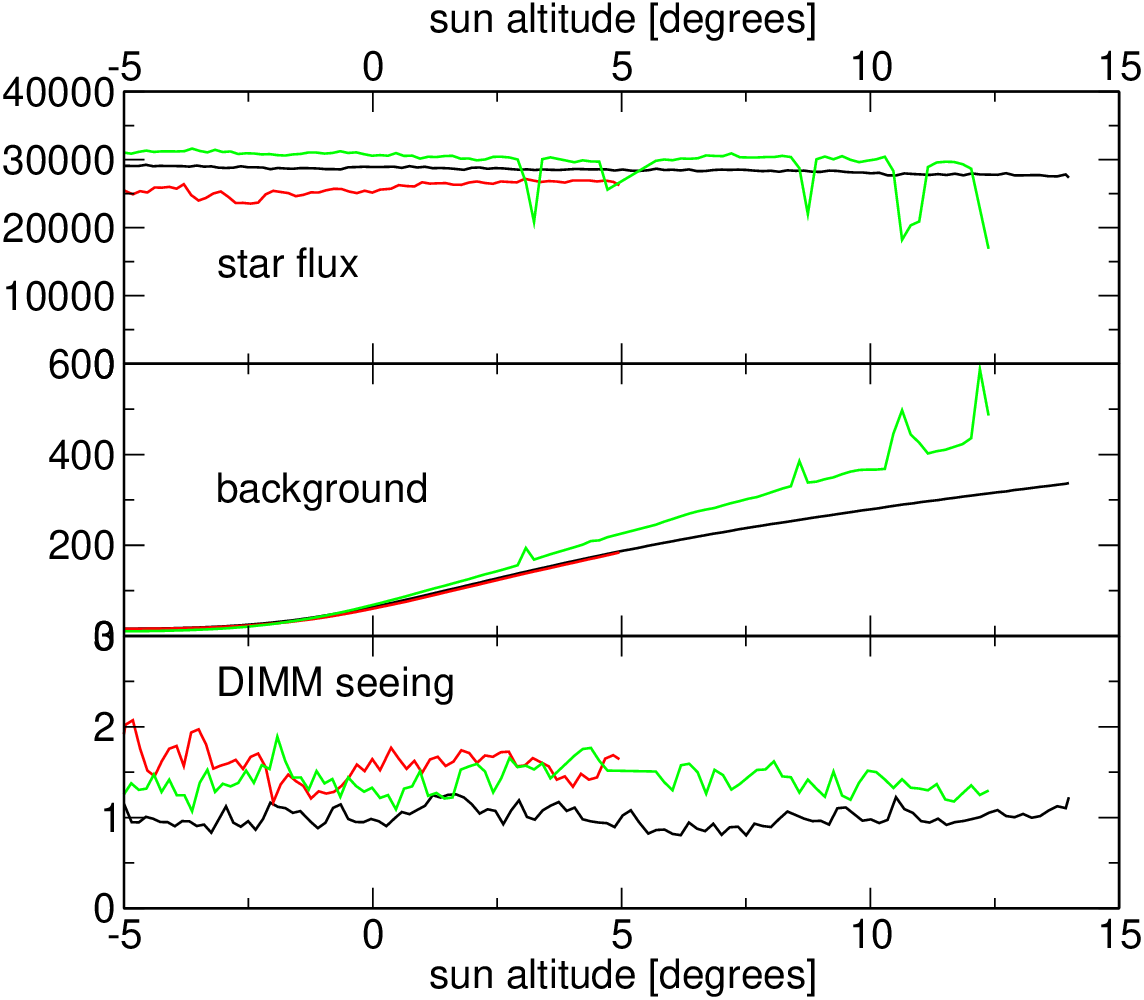}
	\end{tabular}
	\end{center}
	\caption
	{\label{fig:flux_guider_dimm} 
	{\it Left:} Flux in guider; three lines of different colors represent three different experiments. {\it Right:} Flux, background and seeing measured with DIMM (again lines represent different experiments).
	}
\end{figure}

\subsection{Background in guider} \label{sec:method:guider}

Guider was mounted on feeding telescope of MASS-DIMM mainly for pointing because proper pointing accuracy of telescope is about $+1^{\circ}$. This is far greater then DIMM field aperture. Hence we can point to the star if we can detect it in guider. Guider represents simpliest CCD with angular scale of $10^{\prime\prime}/px$. Figure \ref{fig:flux_guider_dimm}, left illustrates dependence of star flux in guider on sun altitude. It can be seen that star disappears in background at sun altitudes $-1^{\circ}$. Relatively high sensitivity to background is caused by low scale.

\subsection{Background in DIMM} \label{sec:method:dimm}

In case of DIMM situation is much better, scale is $0.613^{\prime\prime}/px$. On figure \ref{fig:flux_guider_dimm}, left there are star flux, background level and DIMM seeing for 3 experiments. Experiment indicated by green curves is of particular interest. One can see drops of flux accompanied by increases of background what can be explained by clouds passings. Also there are no any changes in seeing at these moments. This proves that DIMM senstivity to background is low. Our experience shows that we can observe at sun altitudes up to $+13^{\circ}$. Also note that there are no any prominent changes in seeing.

\subsection{Background in MASS} \label{sec:method:mass}

Now I proceed to MASS. MASS is very sensitive to background due to very large stop --- $220^{\prime\prime}$ in diameter. On figure \ref{fig:flux_mass_sunalt_distr}, left\footnote{image from Kornilov, V. Background filtering and approximation in processing of MASS data, see http://curl.sai.msu.ru/mass/download/doc/back\_filter.pdf (in russian)} one can see dependence of background level on sun altitude. Background starts to increase at sun altitude of $-10^{\circ}$ and changes relatively fast so there is need to measure it as frequently as possible --- in our case every 4 minutes. Between two measurements background is interpolated using this empiric law. This approximation works reasonably good until sun altitude of $-6^{\circ}$. We chose this value as treshold for MASS observations.

\section{FEATURES OF OPTICAL TURBULENCE IN TRANSITION PERIOD} \label{sec:ot}

\begin{figure}
	\begin{center}
	\begin{tabular}{cc}
	\includegraphics[height=6.8cm]{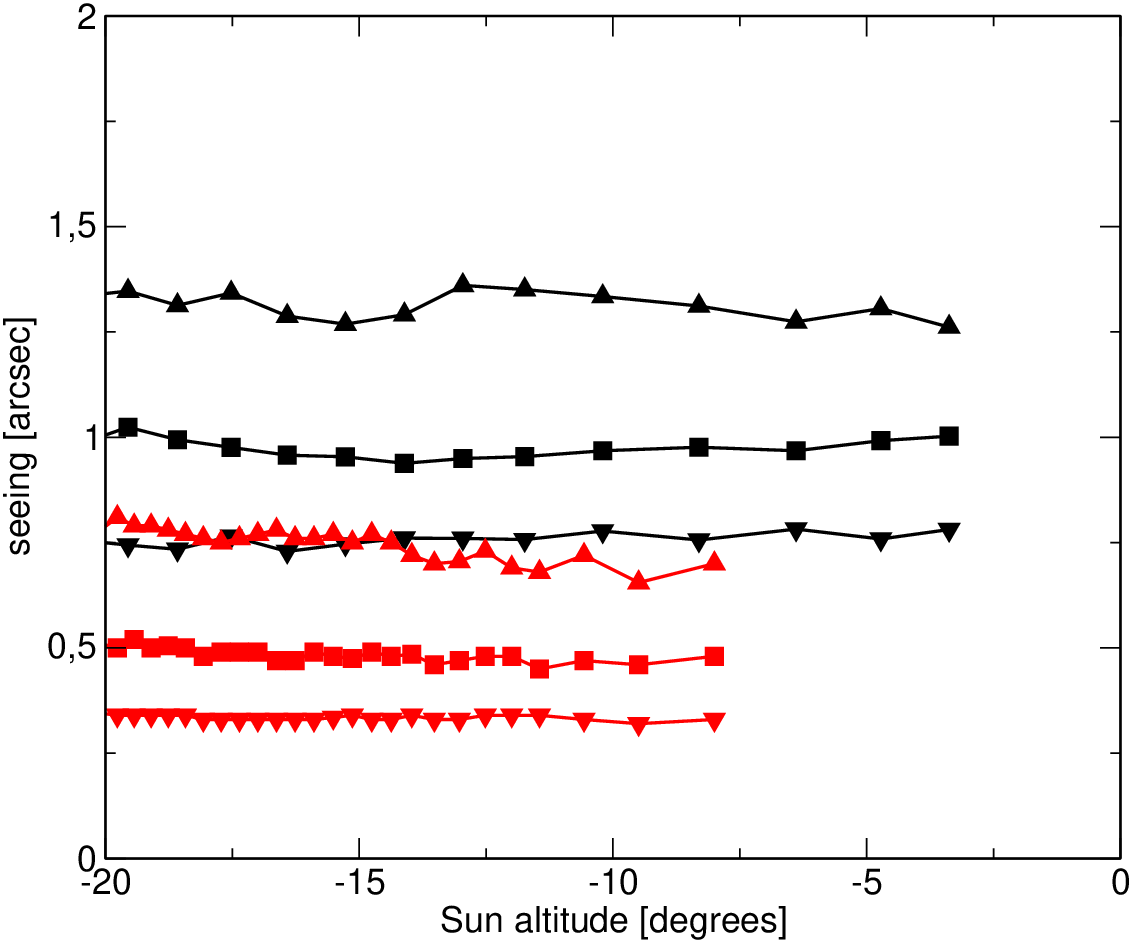}
	\includegraphics[height=6.8cm]{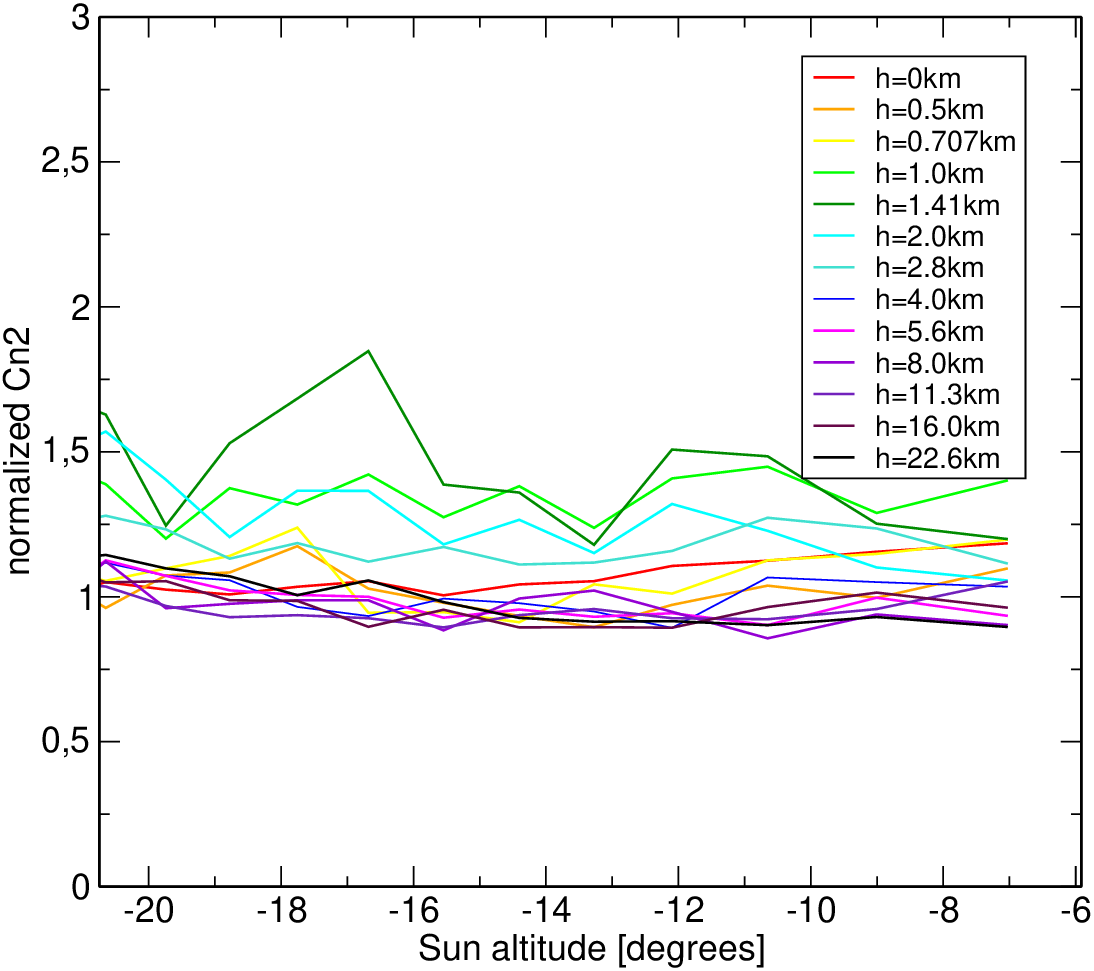}
	\end{tabular}
	\end{center}
	\caption
	{\label{fig:run_med}
	{\it Left:} running medians and quartiles of MASS (red) \& DIMM (black) seeings (lag=1000). Squares stand for the median, triangle down for the 1st quartile, triangle up for the 3rd quartile. {\it Right:} normalized running averages (lag=1000) of turbulence layers intensities.
	}
\end{figure}

\begin{figure}
	\begin{center}
	\begin{tabular}{cc}
	\includegraphics[height=5.4cm]{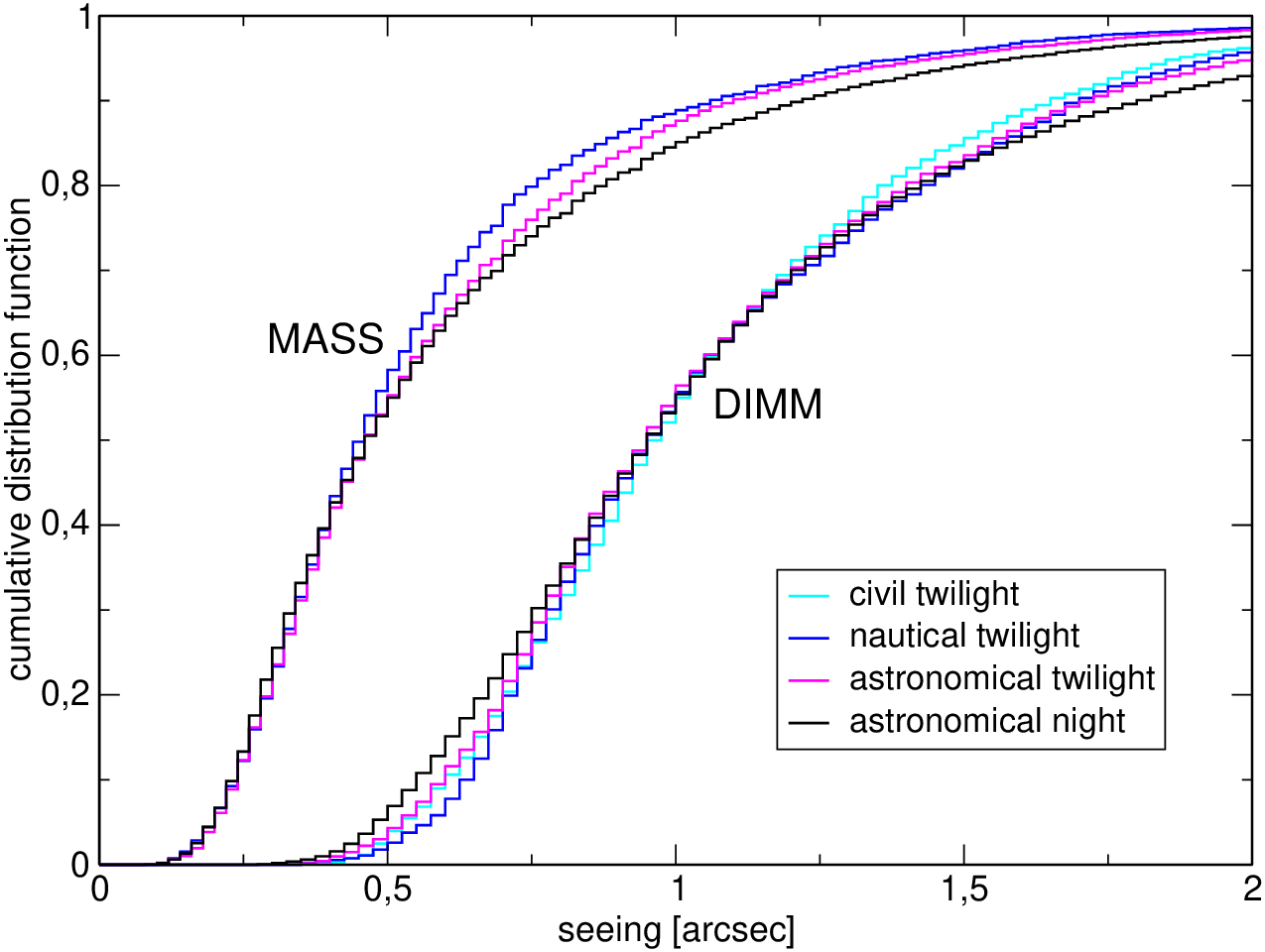}
	\includegraphics[height=5.4cm]{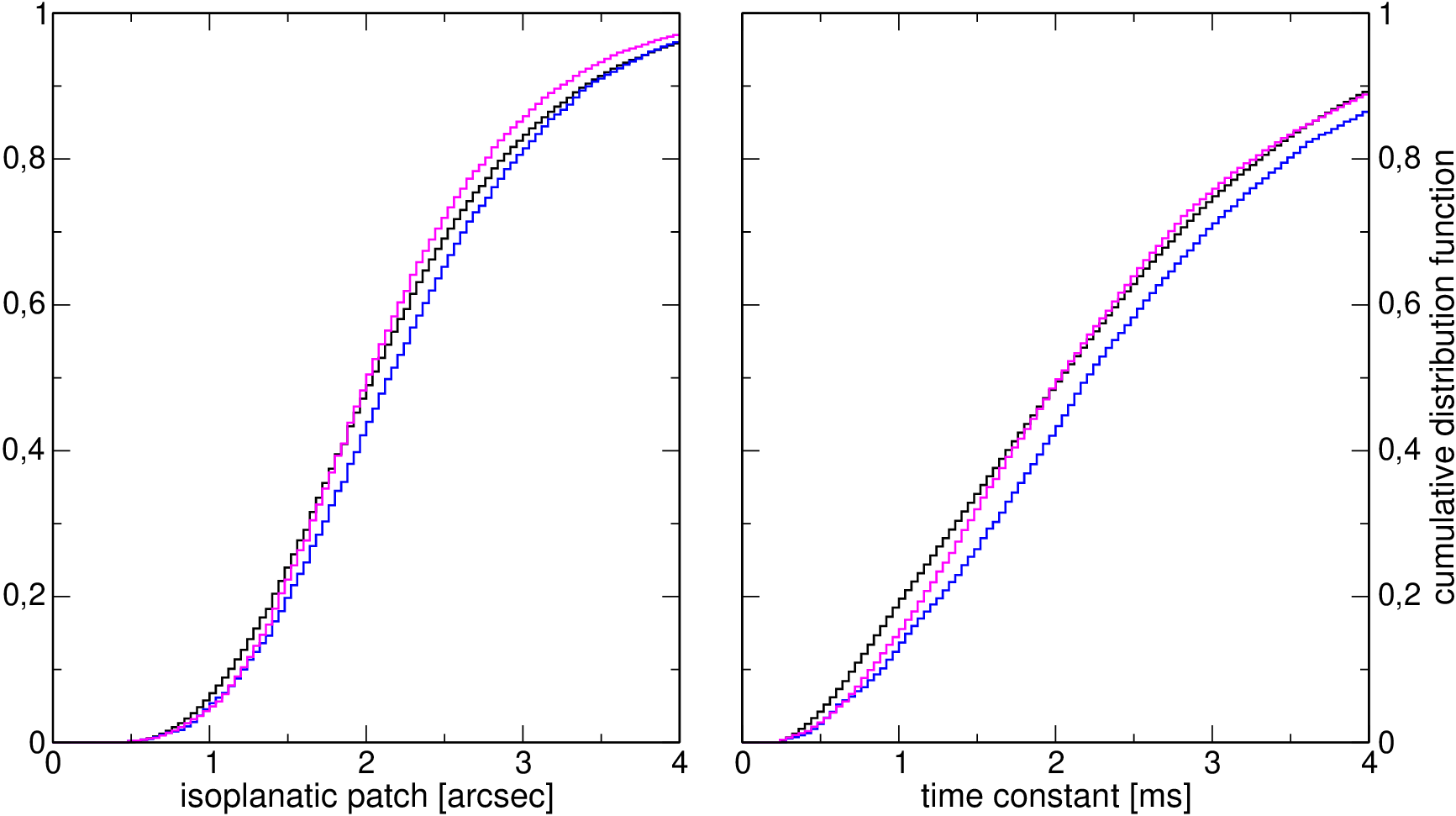}
	\end{tabular}
	\end{center}
	\caption
	{\label{fig:twilight_distrs}
	{\it Left: }DIMM \& MASS seeing distributions. {\it Center:} isoplanatic patch distribution. {\it Right:} time constant distribution.
	}
\end{figure}

On figure \ref{fig:run_med}, left one can see medians and quartiles of MASS \& DIMM seeings computed for 1$^{\circ}$ sun altitude bins. There are no any significant variations in seeings. Distributions of OT parameters (DIMM seeing, MASS seeing, isoplanatic patch, time constant) computed different twilight: civil, nautical, astronomical for also don't reveal any dependence on sun altitude (see fig.\ref{fig:twilight_distrs}).

Also we computed running medians of all 13 turbulent layers normalized by medians evaluated from all dataset (see fig.\ref{fig:run_med}, right). One can see great amount of waves but they are all caused by effects of sampling. Proceeding from this we can conclude that condition of optical turbulence above Shatdjatmaz during night and twilight doesn't differ. However possibly some "switching" occurs at greater sun altitudes.

\section{PREDICTION OF SEEING} \label{sec:prediction}

From the practical point of view twilight observations can be used in prediction of seeing for the forthcoming night. To evaluate its prediction capabilities let's consider the following scheme (suppose that we want to predict the DIMM seeing): 

\begin{enumerate}
\item Observations on big telescope start at sun altitude $-18^{\circ}$.
\item But we want to know the seeing for the forthcoming night in advance, at sun altitude $-12^{\circ}$, to prepare telescope for adequate program.
\item Then we average DIMM seeing measured in interval of sun altitudes from 0 to $-12^{\circ}$. Let's denote respective value $\beta_T$. 
\item Then we divide all nights in 4 classes depending on $\beta_T$. If $\beta_T$ is lesser than 1st quartile of distribution of DIMM seeing than we assign this night class A. If $\beta_T$ is between 1st quartile and median, then night is of class B and so on (see table \ref{tab:classes}). Now one can compare nigths of different classes by comparing distributions of seeing.
\end{enumerate}

% \begin{figure}[b]
% 	\begin{center}
% 	\includegraphics[height=8cm]{classif.eps}
% 	\end{center}
% 	\caption
% 	{\label{fig:classif}
% 	classified
% 	}
% \end{figure}

\begin{wraptable}{l}{80pt}
\caption{\label{tab:classes}
Classes (see the text)}
\begin{center}
\begin{tabular}{r|l}
class & $\beta_T$ \\
\hline
A & $(0,q_1]$ \\
B & $(q_1,q_2]$ \\
C & $(q_2,q_3]$ \\
D & $(q_3,\infty)$ \\
\end{tabular}
\end{center}
\end{wraptable}

\begin{figure}
	\begin{center}
	\includegraphics[height=6.2cm]{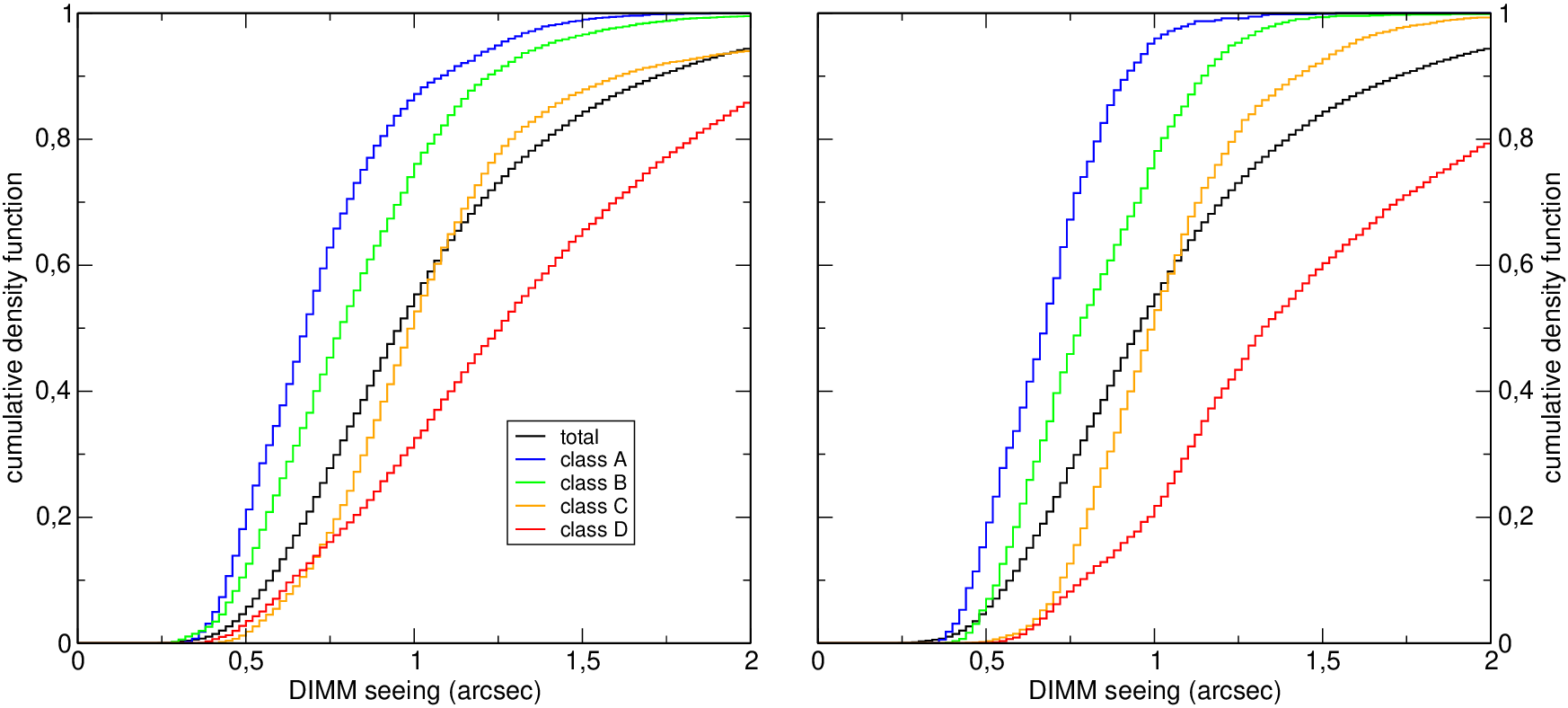}
	\end{center}
	\caption
	{\label{fig:dimm_pred}
	Distribution of DIMM seeing for nights of different classes (see text). {\it Left:} all night, {\it right:} first 3 hours.
	}
\end{figure}

\begin{figure}
	\begin{center}
	\includegraphics[height=6.2cm]{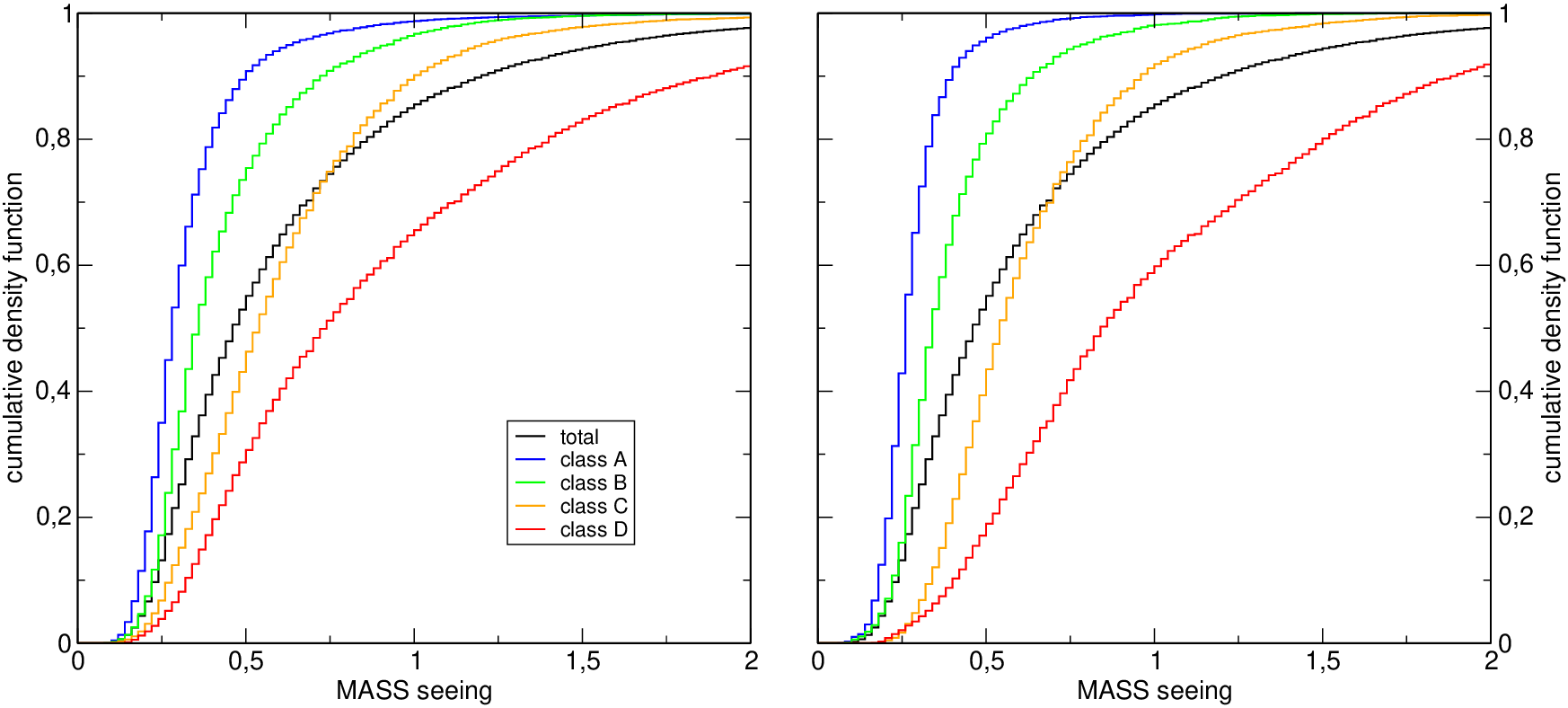}
	\end{center}
	\caption
	{\label{fig:mass_pred}
	Distribution of MASS seeing for nights of different classes (see text). {\it Left:} all night, {\it right:} first 3 hours.
	}
\end{figure}

These distributions is represented on the figure \ref{fig:dimm_pred}. On the left panel there is the all-night seeing and on right panel the first 3 hours seeing. You can see that if in twilight the seeing was good there is a high probability that it will be good for all night and vice versa. For example for the class A night median of DIMM seeing distribution is $0.69^{\prime\prime}$ what is significantly lower than the total median. The effect is slightly more visible for the MASS seeing (fig. \ref{fig:mass_pred}). 

\begin{figure}
	\begin{center}
	\includegraphics[height=10cm]{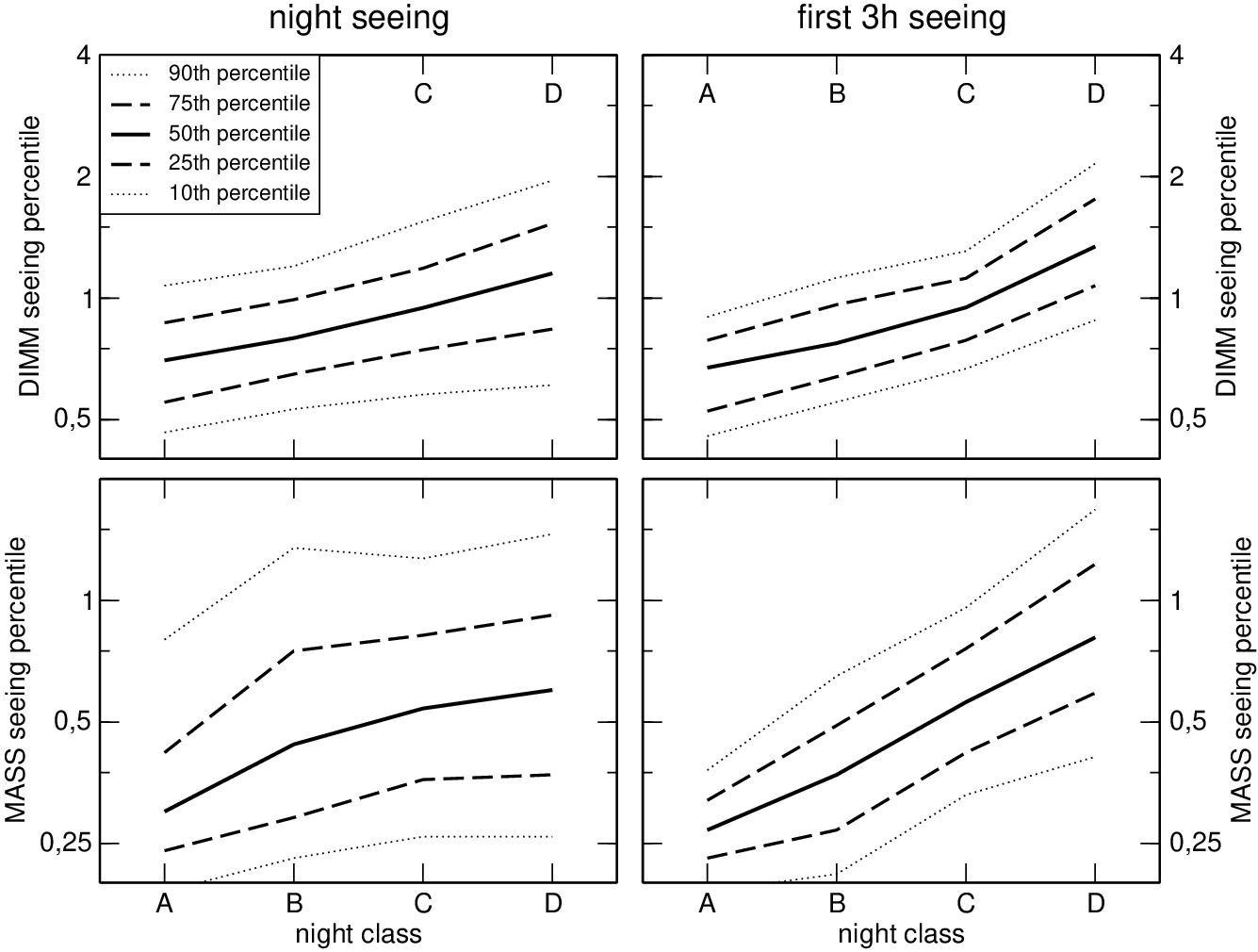}
	\end{center}
	\caption
	{\label{fig:MASSDIMM_percentiles}
	Night class is along x axis, percentiles of distributions on figures \ref{fig:dimm_pred},\ref{fig:mass_pred} along y axis. Dotted lines stand for 10th and 90th percentiles, dashed lines for 25th and 75th percentiles, and solid line for median. Scale along y axis is the same for all graphs.
	}
\end{figure}

Another way to analyse the shape of these distributions is to plot percentiles of seeing distributions for different classes (fig. \ref{fig:MASSDIMM_percentiles}). One can note interesting feature: seeing prediction works better for the MASS seeing than for the overall DIMM seeing. In principle this is expectable --- free atmosphere changes slower than ground layer because of larger scales. This point is closely related to temporal characteristics of seeing discussed by Travouillon, 2008\cite{Travouillon08} and can be considered as another way to describe the seeing variability because it can be applied not only to twilight observations but to any moment during night.

\begin{figure}
	\begin{center}
	\includegraphics[height=6.2cm]{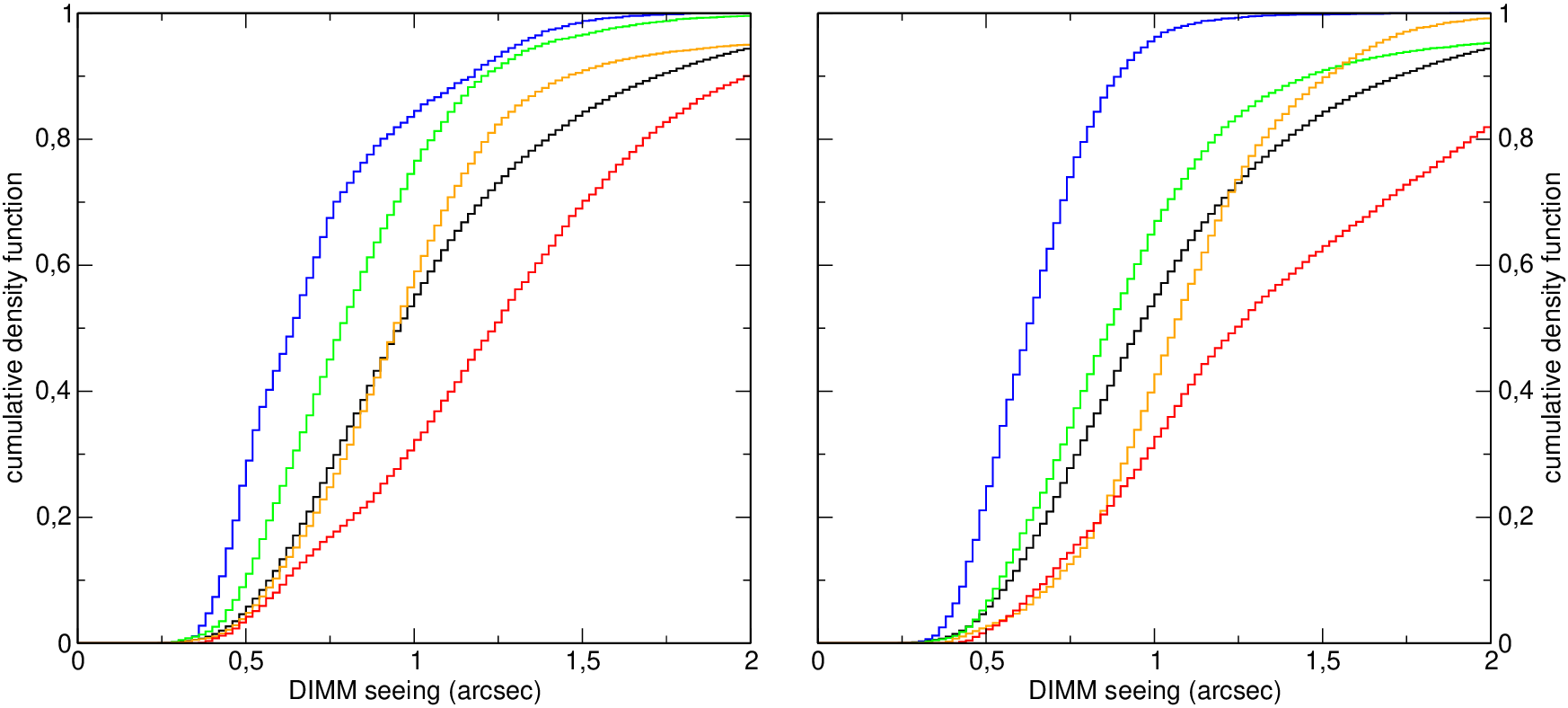}
	\end{center}
	\caption
	{\label{fig:longshort_pred}
	Distribution of DIMM seeing for nights of different classes (see text). {\it Left:} long preparation time, {\it right:} short preparation time.
	}
\end{figure}

All these graphs are valid in case of long preparation time. Now let's imagine that we can switch instrument in a couple of minutes --- this is case of short preparation time. In this case most recent measurements with MASS-DIMM will be more informative --- see fig. \ref{fig:longshort_pred}. On the left panel there is distribution for nights classified by seeing averaged over all twilight $-18^{\circ}<h_{\odot}<0^{\circ}$, on the right panel for nights classified by seeing averaged in $-18^{\circ}<h_{\odot}<-12^{\circ}$. One can see that in second case prediction is better. So we can make quite obvious conclusion: if preparation time is short then there is no sense in observing in twilight.

\newpage
\section{CONCLUSIONS} \label{sec:conclusions}

List of main conclusions of this work:
\begin{enumerate}
\item It's possible to observe in twilight with our particular MASS-DIMM device. Following tresholds of sun altitude for observations with different devices was set.
	\begin{itemize}
	\item Guider $h_{\odot}<0^{\circ}$
	\item DIMM $h_{\odot}<+10^{\circ}$
	\item MASS $h_{\odot}<-6^{\circ}$
	\end{itemize}
\item It seems that there were no significant changes in OT characteristics for investigated interval of sun altitudes.
\item Twilight observations with MASS-DIMM can provide some guess about night conditions of OT in some time before night beginning. This time can be used for program switching. If program switching takes a short time - for example 10 minutes - then it's more reasonable to use most recent seeing values. In this case there is no sense in twilight observations. 
\item Also twilight measurements can be useful if one have possible observational program that require very good seeing and doesn't require dark sky. In this case one can observe in twilight if it's known that seeing is good what is impossible without simultaneous observations with MASS-DIMM.
\end{enumerate}

I have to note that all these conclusions are valid for Mt. Shatdzatmaz. This site deviates from typical desert astronomical site and characteristics of turbulence here might be different.

What we can do to reduce effect of background on:
\begin{itemize}
\item 
MASS
	\begin{itemize}
	\item Reduce jitter of the mount.
	\item Shrink MASS diaphragm to $20^{\prime\prime}$.
	\end{itemize}
\item
Guider
	\begin{itemize}
	\item Improve mount pointing accuracy.
	\item Remove the guider.
	\end{itemize}
\end{itemize}

All these measures are directed to reducing effect of background. In case of MASS reduction of stop size might be very useful. But this is impossible without reduction of jitter of telescope mount. Considering guider: improvement of pointing accuracy might make guider useless. So capability to observe in twilight depends mostly on telescope mount quality.

\acknowledgments

This investigation made use of the data obtained with the SAI ASM constructed by the SAI site testing team: V. Kornilov, N. Shatsky, O. Voziakova, S. Potanin, M. Kornilov. I also would like to give special thanks to my supervisor V. Kornilov for the discussion and help with software development.

%%%%% References %%%%%

\bibliography{report}   %>>>> bibliography data in report.bib
\bibliographystyle{spiebib}   %>>>> makes bibtex use spiebib.bstl

\end{document}